\documentclass[12pt]{iopart}

\usepackage{iopams}
\expandafter\let\csname equation*\endcsname\relax
\expandafter\let\csname endequation*\endcsname\relax
\usepackage{amsmath}
\usepackage[english]{babel}
\usepackage{graphicx}
\usepackage{cite}
\usepackage{color}
\usepackage{bbm}
\usepackage[normalem]{ulem}
\usepackage{times}
\usepackage[colorlinks=true,citecolor=cyan,linkcolor=cyan,urlcolor=cyan]{hyperref}
\usepackage[normalem]{ulem}
\usepackage{dsfont}
\usepackage{txfonts}

\newcommand{\ket}[1]{\vert #1 \rangle}
\newcommand{\bra}[1]{\langle #1 \vert}
\newcommand{\ketbra}[2]{\vert #1 \rangle \langle #2 \vert}

\usepackage{mathptmx} 

\begin{document}

\title{Dynamical quantum phase transitions and non-Markovian dynamics} 

\author{Thi Ha Kyaw}
\address{Centre for Quantum Technologies, National University of Singapore, 3 Science Drive 2, Singapore 117543, Singapore}
\author{Victor M. Bastidas}
\address{NTT Basic Research Laboratories \& Research Center for Theoretical Quantum Physics,  3-1 Morinosato-Wakamiya, Atsugi, Kanagawa, 243-0198, Japan}
\author{Jirawat Tangpanitanon}
\address{Centre for Quantum Technologies, National University of Singapore, 3 Science Drive 2, Singapore 117543, Singapore}
\author{Guillermo Romero}
\address{Departamento de F\'isica, Universidad de Santiago de Chile (USACH), Avenida Ecuador 3493, 9170124, Santiago, Chile}
\address{Center for the Development of Nanoscience and Nanotechnology, Estaci\'on Central, 9170124, Santiago, Chile}
\author{Leong-Chuan Kwek}
\address{Centre for Quantum Technologies, National University of Singapore, 3 Science Drive 2, Singapore 117543, Singapore}
\address{MajuLab, CNRS-UNS-NUS-NTU International Joint Research Unit, UMI 3654, Singapore}
\address{National Institute of Education, Nanyang Technological University, 1 Nanyang Walk, Singapore 637616, Singapore}

\date{\today}

\begin{abstract}
In the context of closed quantum systems, when a system prepared in its ground state undergoes a sudden quench, the resulting Loschmidt echo can exhibit zeros, resembling the Fisher zeros in the theory of classical equilibrium phase transitions. These zeros lead to nonanalytical behavior of the corresponding rate function, which is referred to as \textit{dynamical quantum phase transitions} (DQPTs).
In this work, we investigate DQPTs in the context of open quantum systems that are coupled to both Markovian and non-Markovian dephasing baths via a conserved quantity. The general framework is corroborated by studying the non-equilibrium dynamics of a transverse-field Ising ring. We show the robustness of DQPT signatures under the action of both engineered dephasing baths, independently on how strongly they couple to the quantum system. Our theory provides insight on the effect of non-Markovian environments on DQPTs. 
\end{abstract}

\maketitle

\section{Introduction}\label{sec:intro}
The theory of equilibrium phase transitions is well-studied in statistical mechanics and thermodynamics, and it provides us with an excellent framework to understand and characterize phases of matter at  zero temperature~\cite{sachdev2007,suzuki2012}.
In the theory of classical phase transitions, nonanalytical behavior can appear in thermodynamic potentials during a phase transition. This is related to Lee-Yang zeros of the grand canonical potential \cite{yang1952statistical,lee1952statistical} or Fisher zeros of the canonical one \cite{fisher1965lectures}, even with a perfectly well-defined microscopic Hamiltonian without any singular interactions. To illustrate the origin of the aforementioned nonanalytical behavior, let us consider the free energy density
$
	f =-\lim_{N\rightarrow \infty} (N\beta)^{-1}\log [Z(\beta)]
	\ ,
$
where $N$ is the number of particles, $\beta$ is the inverse temperature, and $Z(\beta)=\textrm{Tr}(e^{-\beta {H}})$ is the canonical partition function. From this, one can see that non-analytical behavior in macroscopic quantities such as the free energy density occurs whenever the partition function $Z(\beta)$ becomes zero.

In recent years, there have been many attempts to generalize the concept of Lee-Yang zeros to nonequilibrium quantum dynamics~\cite{pollmann2010dynamics}. M. Heyl et. al.~\cite{heyl2013dynamical}, suggested that there is a dynamical counterpart of equilibrium quantum phase transitions (QPTs), referred to as dynamical quantum phase transitions (DQPTs). In fact, the concept of DQPTs is intimately related to quantum quenches in many-body systems. Let us consider a quantum many-body Hamiltonian $\hat{H}(\lambda)$ with a quantum critical point at $\lambda=\lambda_c$. Here, $\lambda$ is an external control parameter. DQPTs may be observed when a quantum system undergoes a sudden quench from $\hat{H}(\lambda_i)$ to $\hat{H}(\lambda_f)$, where $\lambda_i$  and $\lambda_f$ are the initial and final control parameters. Depending on nature of the quench, there are two classes of DQPTs \cite{vzunkovivc2018dynamical,Halimeh2,Halimeh3,Halimeh4,lang2018concurrence,Halimeh2018}. The first class, DPT-I \cite{moeckel2008interaction,moeckel2010crossover,sciolla2010quantum,sciolla2011dynamical,gambassi2011quantum,sciolla2013quantum,maraga2015aging,mori2018thermalization,zhang2017,Halimeh1}, describes a type of dynamical phase transition in which a time-averaged order parameter is non-zero in the long-time limit for quenches $\lambda_f<\lambda_c$, but vanishes for quenches across the critical point $\lambda_c$, i.e., $\lambda_f >\lambda_c > \lambda_i$. The second class, DPT-II \cite{heyl2013dynamical,karrasch2013dynamical,andraschko2014dynamical,heyl2014dynamical,jurcevic2017direct,heyl2018dynamical,PIROLI2018454,PhysRevLett.121.130603}, generalizes the notion of nonanalyticity in the free energy density to the nonequilibrium dynamics, based on the complex Loschmidt amplitude (along the paper, we set $\hbar=1$)
\begin{equation}
	\mathcal{G}(t)=\bra{\psi(0)}e^{-i\hat{H}(\lambda_f)t}\ket{\psi(0)}.\label{Loschmidt_amp}
\end{equation}
Here, the initial state $\ket{\psi(0)}=\ket{E_0(\lambda_i)}$ is the ground state of the initial Hamiltonian, i.e., $\hat{H}(\lambda_i)\ket{E_0(\lambda_i)}=E_0(\lambda_i)\ket{E_0(\lambda_i)}$. Such a ground state can represent a quantum phase of matter. To investigate DQPTs, we let the system evolve under a new Hamiltonian $\hat{H}(\lambda_f)$ and study the dynamics of the return probability. The function $\mathcal{G}(t)$ resembles the partition function $Z(\beta)$ in statistical mechanics. The return rate associated to the Loschmidt amplitude $\mathcal{G}(t)$ is defined as
\begin{equation}
     \label{eq:Ratefunction}
\zeta(t) = -\lim_{N\rightarrow \infty} \frac{1}{N}\log[\mathcal{G}(t)]
\end{equation}
resembling the free energy in the context of classical phase transitions. In analogy to the Fisher zeros in the equilibrium statistical mechanics, one can study the zeros of the Loschmidt echo $L(t)=|\mathcal{G}(t)|^2$. The latter would lead to singular behavior of the return rate $\zeta(t)$, which has been extensively studied in the context of nonequilibrium quantum phase transitions~\cite{quan2006decay,li2007density,wei2012lee,dora2013loschmidt,wei2014phase,suzuki2016dynamics,Choo2017}. In the case of the one-dimensional quantum Ising model, it has been shown that Fisher zeros of $L(t)$ form lines in the complex plane, touching the real axis only for a quench across the equilibrium critical point \cite{heyl2013dynamical}. Thus, the zeros in $\mathcal{G}(t)$ at critical times $t_c$ lead to non-analyticities in $\zeta(t)$ \cite{heyl2013dynamical,karrasch2013dynamical,andraschko2014dynamical,heyl2014dynamical,zvyagin2016,jurcevic2017direct,heyl2018dynamical}, whenever the evolved state $\ket{\psi(t)}$ becomes orthogonal to the initial one $\ket{\psi(0)}$. 
Although DPT-I and DPT-II seem to have different origins, numerical investigations suggest that in the presence of sufficiently long-range interactions at zero temperature, they are intimately related~\cite{vzunkovivc2018dynamical}. Throughout the paper, we will focus on the DPT-II class.

Symmetry plays a fundamental role in quantum phase transitions. For example, in the thermodynamic limit, the ground state of a many-body system can exhibit less symmetries than the system Hamiltonian, a phenomenon known as spontaneous symmetry breaking. In the case of DPT-II, we can observe a similar symmetry breaking mechanism that can be captured by the Loschmidt amplitude. In particular, let us consider $M$-fold degenerate ground states $\ket{E^{(j)}_0(\lambda_i)}$ of the initial Hamiltonian $\hat{H}(\lambda_i)$, where $j=1,\cdots,M$. This degeneracy appears due to the $\mathbb{Z}_M$ symmetry of the underlying Hamiltonian. By initializing the system in one of the ground states $\ket{\psi(0)}=\ket{E^{(l)}_0(\lambda_i)}$, one can generalize the Loschmidt amplitude by considering the probability to remain in the degenerate ground-state manifold \cite{heyl2014dynamical,vzunkovivc2018dynamical}, as follows
\begin{equation}
	L_{\text{Sym}}(t)=\sum_{j=1} ^M |\bra{E^{(j)}_0(\lambda_i)}e^{-i\hat{H}(\lambda_f)t}\ket{\psi(0)}|^2 \label{prob_ground_state}
	\ .
\end{equation}
Here the individual terms of the sum decay exponentially with the system size $N$, i.e., $|\bra{E^{(j)}_0(\lambda_i)}e^{-i\hat{H}(\lambda_f)t}\ket{\psi (0)}|^2 =\exp[-N (\zeta_j(t)+\zeta_j ^\ast (t))]$. In the thermodynamic limit $N\rightarrow \infty$, only one of them dominates such that $L_{\text{Sym}}(t)=\exp[-N {S}(t)]$, where $S(t)=\textrm{min}_j \zeta_j(t)$. In this way, the $\mathbb{Z}_M$ symmetry, broken by the initial configuration, is restored at the critical times when all $\zeta_j$ are equal~\cite{vzunkovivc2018dynamical}, leading to a cusp in $S(t)$ \cite{andraschko2014dynamical,jurcevic2017direct}. In this case, one can interpret DQPTs in terms of the dynamical restoration of symmetry \cite{heyl2014dynamical}, rather than the orthogonality between the initial state and the time-evolved state.

Strictly speaking, a non-analyticity of the return rate occurs when the initial state becomes orthogonal to the evolved state after the sudden quench. In addition, it is not yet settled under which general circumstances such condition occurs \cite{jurcevic2017direct,heyl2017dynamical}. However, one necessary condition seems to be the existence of a sufficiently strong quench, which is achieved when a system parameter $\lambda$ is quenched across an underlying equilibrium critical point $\lambda_c$. Recent experiments in trapped ions \cite{jurcevic2017direct, zhang2017}, ultracold atoms in optical lattices \cite{flaschner2016}, and quantum simulation by using superconducting qubits~\cite{guo2018} seem to support this feature. Nevertheless, it is also interesting to note that there is no one-to-one correspondence between DQPTs and equilibrium QPTs. It is possible to have non-analytical behavior in the return rate without crossing an equilibrium critical point $\lambda_c$ in the quench, and one can cross a critical point during a quench without divergencies in the return rate \cite{vajna2015topological}. In addition, there is a recent attempt to relate DQPTs and equilibrium QPTs near the vicinity of DQPT in the transverse-field Ising chain \cite{trapin2018constructing}.

\begin{figure}[t]
\centering
\includegraphics[scale=0.28]{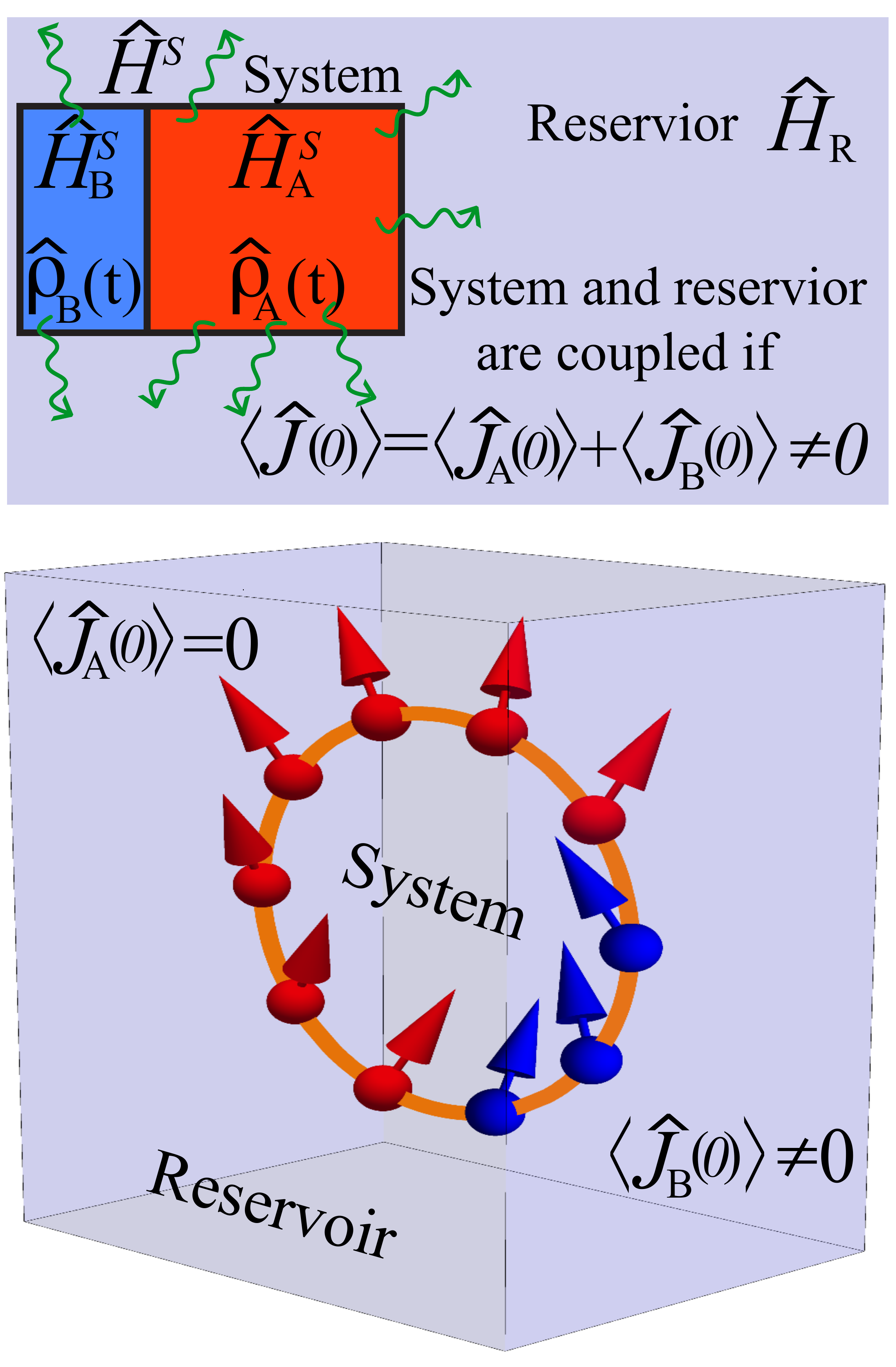}
\caption{Open quantum system approach to DQPTs. The system of interest with Hamiltonian $\hat{H}^{S}$ is composed of two transverse Ising chains A and B depicted in red and blue, respectively. The spin chain A is initialized in one of the ground states of the Hamiltonian $\hat{H}^{S}_{A}$ (see main text) without current $\langle\hat{J}_{A}(0) \rangle =0$, while the spin chain B is initialized in a ground state of the Hamiltonian  $\hat{H}^{S}_{B}$ (see main text),  which plays the role of an energy current source $\langle\hat{J}_{B}(0) \rangle \neq 0$. In the main text we describe in detail how one can create a quantum state with non-vanishing current inside the ring. The entire ring is coupled to the enviroment via a conserved quantity $\hat{J}$ which is the total energy current. The system $\hat{H}^{S}$ is coupled to the environment $\hat{H}_{R}$ when the initial total current is different from zero, i.e., $\langle\hat{J}(0)\rangle=\langle\hat{J}_{A}(0)\rangle  +\langle\hat{J}_{B}(0) \rangle \neq 0$. In the manuscript, $\hat{\rho}_{A/B}$ denote the reduced density matrices of the subsystems A and B.}
\label{fig:closedRing}
\end{figure}

In general, when a quantum systems is in a statistical ensemble, it is convenient to use the density matrix formalism.
Therefore, in order to extend the notion of DQPTs to mixed states, a generalization of the Loschmidt amplitude Eq.~(\ref{Loschmidt_amp}) in terms of density matrices is indispensable. There are two ways to achieve this. Firstly, \textit{the interferometric Loschmidt amplitude} \cite{bhattacharya2017mixed,heyl2017dynamical,lang2018concurrence}
\begin{equation}
	\mathcal{G}_I (t)= \textrm{Tr}[\hat{\rho}(0)\hat{U}(t,0)],
\end{equation}
which is a direct extension of Eq. (\ref{Loschmidt_amp}), where $\hat{\rho}(0)$ is the density matrix at initial time, and $\hat{U}(t,0)$ is the unitary evolution operator from the initial time to later time $t$. This method is useful when a system, initially prepared in a statistical ensemble, undergoes a unitary evolution via a sudden quench without interacting with an external environment. However, the interferometric Loschmidt amplitude $\mathcal{G}_I (t)$ cannot be used when the dynamics are non-unitary, i.e., when the system couples to a reservior. Secondly, \textit{the fidelity Loschmidt amplitude} is defined as
\begin{equation}\label{eq:fidelity_Loschmidt}
	\mathcal{G}_F (t)= \textrm{Tr}\left[\sqrt{\sqrt{\hat{\rho}(0)}\hat{\rho}(t)\sqrt{\hat{\rho}(0)}}\right],
\end{equation}
which is a metric, measuring the distance between the time-evolved density matrix $\hat{\rho}(t)$ and the initial one $\hat{\rho}(0)$. Furthermore, the \textit{fidelity Loschmidt amplitude} \cite{sedlmayr2017fate,mera2018dynamical,lang2018concurrence,Bandyopadhyay:2018aa} and related measures \cite{PhysRevB.98.134310} have proven useful for analyzing DPT-II when including dissipative mechanisms. The advantage of this formulation is that $\hat{\rho}(t)$ can now evolve under both unitary and non-unitary dynamics. In the present work, we adopt the latter quantifier extensively.

This work is inspired by a recent direct experimental observation of DQPTs in a system of six trapped ions~\cite{jurcevic2017direct} that shows indirect signatures \cite{heyl2014dynamical} of DQPTs. Our aim is to understand how dephasing channels acting upon quantum systems affect the fidelity Loschmidt amplitude and signatures of DQPTs. In particular, we consider quenches in the context of open quantum systems that are described by Lindblad-type master equations. Furthermore, we investigate the effect of non-Markovian environments~\cite{bastidas2018} on signatures of DQPTs experienced by an Ising chain. We show that signatures of DQPT are robust under the action of both Markovian and non-Markovian dephasing channels; no matter how strongly they couple to the quantum system. To the best of our knowledge, there are no previous works describing the effect of a non-Markovian environment on DQPTs. Based on an open quantum system approach, we obtain an exact master equation that is valid for any coupling strength between the spin chain and the dephasing bath. Our framework can be applied to other systems experiencing DQPTs under the effect of non-Markovian environments.

This article is organized as follows. First, we introduce a framework to describe a quantum system coupled to both Markovian and non-Markovian baths in Sec. \ref{sec:DQPT_env}. In Sec. \ref{subsec:setup} we apply the general framework to a paradigmatic model in the theory of both QPTs and DQPTs: a one-dimensional transverse Ising model with periodic boundary conditions~\cite{sachdev2007}. In addition, we discuss numerical signatures of DQPTs under Markovian, non-Markovian and both Markovian and non-Markovian dynamical evolutions in Secs. \ref{subsec:signatures} \& \ref{subsec:interplay}. Lastly, in Sec. \ref{sec:conclusion}, we provide concluding remarks and outlook.

\section{Open system approach to dynamical quantum phase transitions}\label{sec:DQPT_env}
It is experimentally challenging, in general, to observe signatures of quantum criticality. One of the reasons is the existence of finite temperature effects that do not allow us to strictly investigate quantum phases of matter at the absolute zero temperature \cite{sachdev2007}. Similarly, it is not a easy task to resolve signatures of DQPTs in experiments. Previously, we have seen that the Lochsmidt echo has the expression
$
L(t)\equiv \exp[{-N(\zeta(t)+\zeta^\ast (t))}],
$
which implies that the return rate is hard to be resolved when the system size increases, especially in the thermodynamic limit $N\rightarrow \infty$, where DQPTs are supposed to occur. However, thanks to \textit{the minimization principle} introduced in Refs. \cite{andraschko2014dynamical,heyl2014dynamical}, an indirect measurement of DQPTs is possible by observing cusps in the rate function $\varpi(t)$ defined below. By using the rate function $\varpi(t)$, one can observe signatures of DQPTs even in a finite-size quantum system, when the initial states breaks the underlying symmetry of the Hamiltonian. The cusps occur whenever the system transits from one symmetry to another, as we have discussed in Sec. \ref{sec:intro} in relation to Eq. (\ref{prob_ground_state}). At the critical times $t_c$, the cusps indirectly indicate zeros in the Lochsmidt amplitude. They become pronounced in the thermodynamic limit \cite{heyl2014dynamical}.

\subsection{The setup}\label{subsec:setup}

The system we consider is the Ising ring composed of two transverse Ising chains $A$ and $B$ as it can be seen in Fig.~\ref{fig:closedRing}. The Ising ring Hamiltonian is
\begin{equation}
	\hat{H}^S= -\tau\sum_{j=1}^{N-1}\hat{\sigma}^x _j \hat{\sigma}^x _{j+1}-H\sum_{j=1}^N \hat{\sigma}^z _j ,\label{eq:HamIsing}
\end{equation} 
where $\hat{\sigma}^{z/x}$ are the Pauli matrices and $N=N_A +N_B$ is the sum of the total spins from the two subsystems. In the following, the sites $j=1,\ldots, N_A$ label the Ising chain A and $j=N_{A+1},\ldots, N$ the Ising chain B. 
The Hamiltonian \eqref{eq:HamIsing} has been extensively studied \cite{sachdev2007} and can be solved analytically even for its sudden quench dynamics in certain limits (see Ref. \cite{suzuki2012} and references therein). The transverse Ising model has a quantum critical point $\tau=H$ \cite{sachdev2007} at $0$ K temperature. Furthermore, if the Ising chain is put into a ring geometry with the periodic boundary condition, $\hat{\sigma}_{N+1}=\hat{\sigma}_1$, as shown in Fig.~\ref{fig:closedRing}, a conserved quantity known as the global current $\hat{J}$ naturally arises \cite{antal1997,wu2008energy}, i.e., $[\hat{H}^S , \hat{J}]=0$ (see \ref{appen:global_current} for its derivation). The expression for the conserved current is
\begin{equation}\label{global_current}
	\hat{J} = \frac{H\tau}{2}\sum_{j} \hat{\sigma}^y _j \left(\hat{\sigma}^x _{j-1}- \hat{\sigma}^x _{j+1} \right).
\end{equation} 
\begin{figure}[t]
\centering
\includegraphics[scale=0.15]{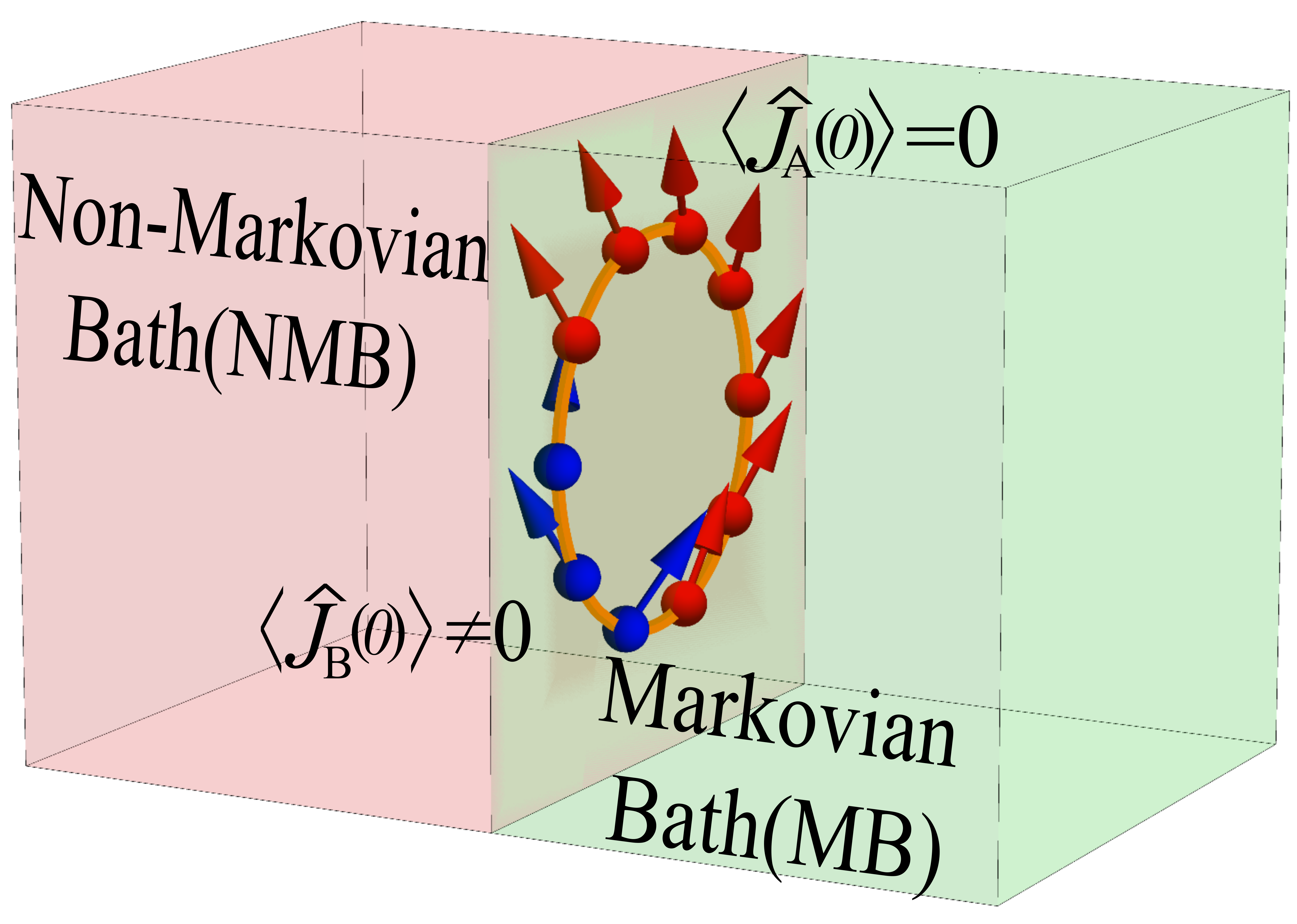}
\caption{The same Ising spin ring seen in Fig. \ref{fig:closedRing}, is globally coupled with both Markovian and non-Markovian dephasing channels. The baths degree of freedom are captured via quantum bosonic modes. Here, the two baths do not interact with each other. }
\label{fig:spinRing_baths}
\end{figure}
One can define local currents $\hat{J}_A$ and $\hat{J}_B$ for the Ising chains A and B, respectively. In contrast to the global current $\hat{J}$, the local currents are not conserved quantities.
One can obtain its explicit expressions by using Eq.~\eqref{global_current} with $j=1,\ldots, N_A$ for the Ising chain A and $j=N_{A}+1,\ldots, N$ for the Ising chain B.
As we have mentioned earlier, DQPT is observed when a quantum system undergoes a sudden quench across equilibrium critical point. In order to apply the minimization principle \cite{jurcevic2017direct, heyl2014dynamical} within the open quantum systems framework of Floquet stroboscopic divisibility that we proposed in a recent paper~\cite{bastidas2018} (see \ref{appen:general}), we treat the Ising chain $A$ as a drain and $B$ as a current source, which will be clarified below. Then, we consider a particular quench protocol, which involves three steps. First, we initialize the Ising chain $A$ in its one of the degenerate ground states of $\hat{H}^{S}_A=-\tau\sum^{N_A-1}_{j=1}\hat{\sigma}^x _j \hat{\sigma}^x _{j+1}$, since it has broken $\mathbb{Z}_2$ symmetry. In particular, $\ket{\psi_+}=\ket{\rightarrow \rightarrow \cdots \rightarrow}$ \& $\ket{\psi_{-}}=\ket{\leftarrow \leftarrow \cdots \leftarrow}$, where $\sigma^x _i \ket{\psi_+}=\ket{\psi_+}$, as well as $\sigma^x _i \ket{\psi_-}=-\ket{\psi_-}$, $\forall i \in A$. We take $\ket{\psi_+}$ as the initial state for the chain $A$. Secondly, we take the chain $B$ initial state as the ground state $\ket{\psi_G}$ of the following Hamiltonian
\begin{equation}\label{ring_B_Hamiltonian}
		\hat{H}^{S}_B=-\tau\sum^{N-1}_{j=N_A+1}\hat{\sigma}^{x}_{j}\hat{\sigma}^{x}_{j+1} -H \sum_{j=N_A+1}^{N} \hat{\sigma}^z _{j}-\nu \hat{J}_B.
\end{equation}
Here, the expression of $\hat{J}_B$ differs from the global current operator since it only applies to the chain $B$ with open boundary condition. Its boundary terms consist only of $-\hat{\sigma}^y _{N_{A}+1}\hat{\sigma}^x _{N_{A}+2}$ and $\hat{\sigma}^y _{N}\hat{\sigma}^x _{N -1}$. The parameter $\nu$ controls the amount of energy current present inside the ring. In this way, we induce the energy current inside the chain $B$ \cite{antal1997}. By combining the Ising chains A and B one forms a Ising ring with a conserved energy current $\hat{J}$ throughout its evolution. Hence, we refer to the subsystem $A$  as `drain' and the subsystem $B$ as `source'. Now we can define the initial state of the Ising ring as $\ket{\psi(0)}=\ket{\psi_{+}}\otimes\ket{\psi_{G}}$, from which we construct the initial density matrix of the system $\hat{\rho}_S(0)=\ket{\psi(0)}\bra{\psi(0)}=\hat{\rho}_A(0)\otimes\hat{\rho}_B(0)$. 

\begin{figure}[t]
\centering
\includegraphics[scale=0.55]{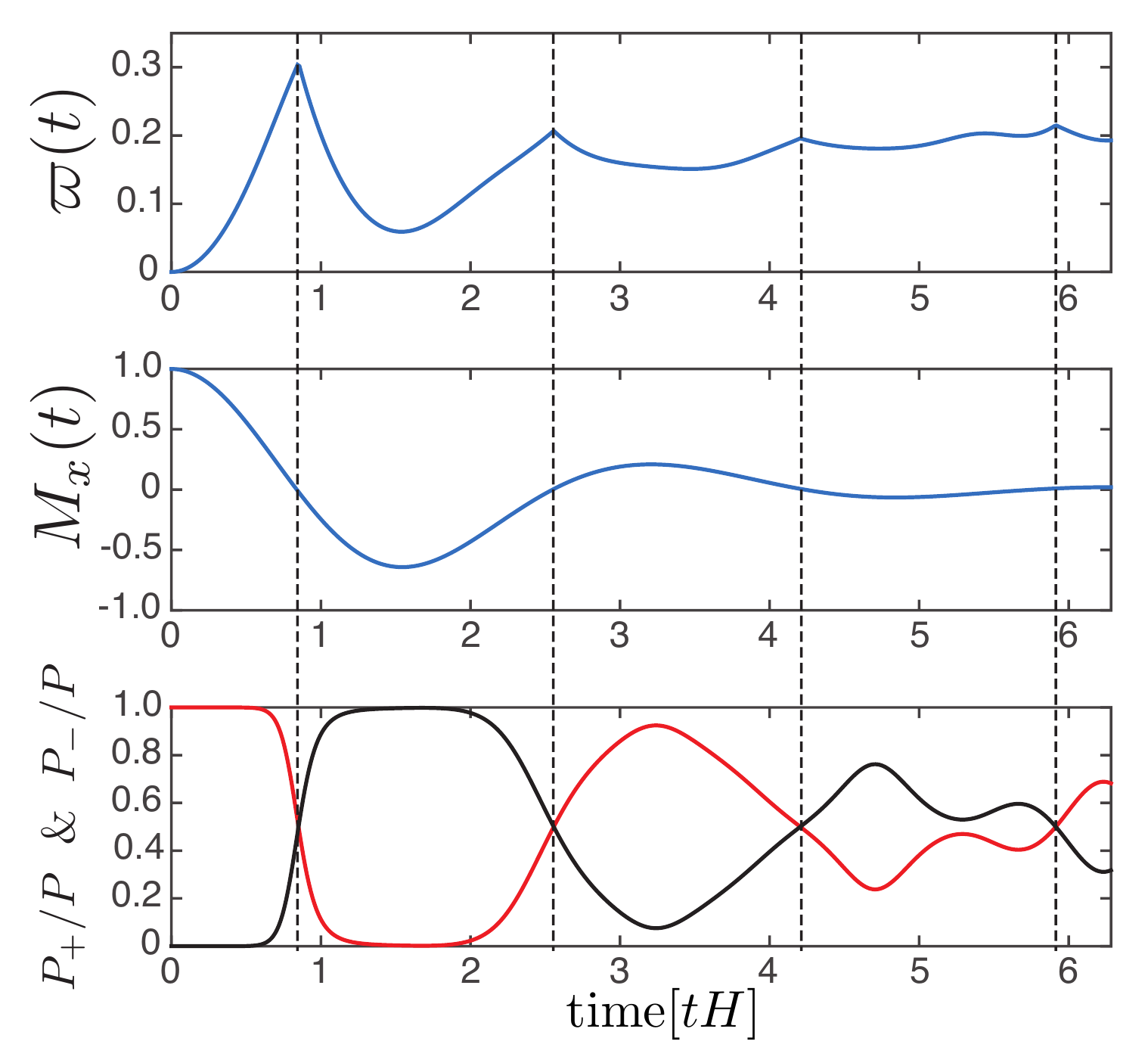}
\caption{Rate function, magnetization, and probabilities are plotted for one period $T$. The system is completely isolated from the baths. Black dash lines represent critical times where the dynamical quantum phase transitions occur. We have $N_A=6$ spins for the subsystem $A$ and $N_B=2$ spins for the subsystem $B$. The interspin coupling is $\tau=0.42\Omega$ and the transverse magnetic field $H=\Omega$, $\nu=5\Omega$ and $\gamma(t)=0$.}
\label{fig:no_bath_collage}
\end{figure}

\begin{figure}[t]
\centering
\includegraphics[scale=0.32]{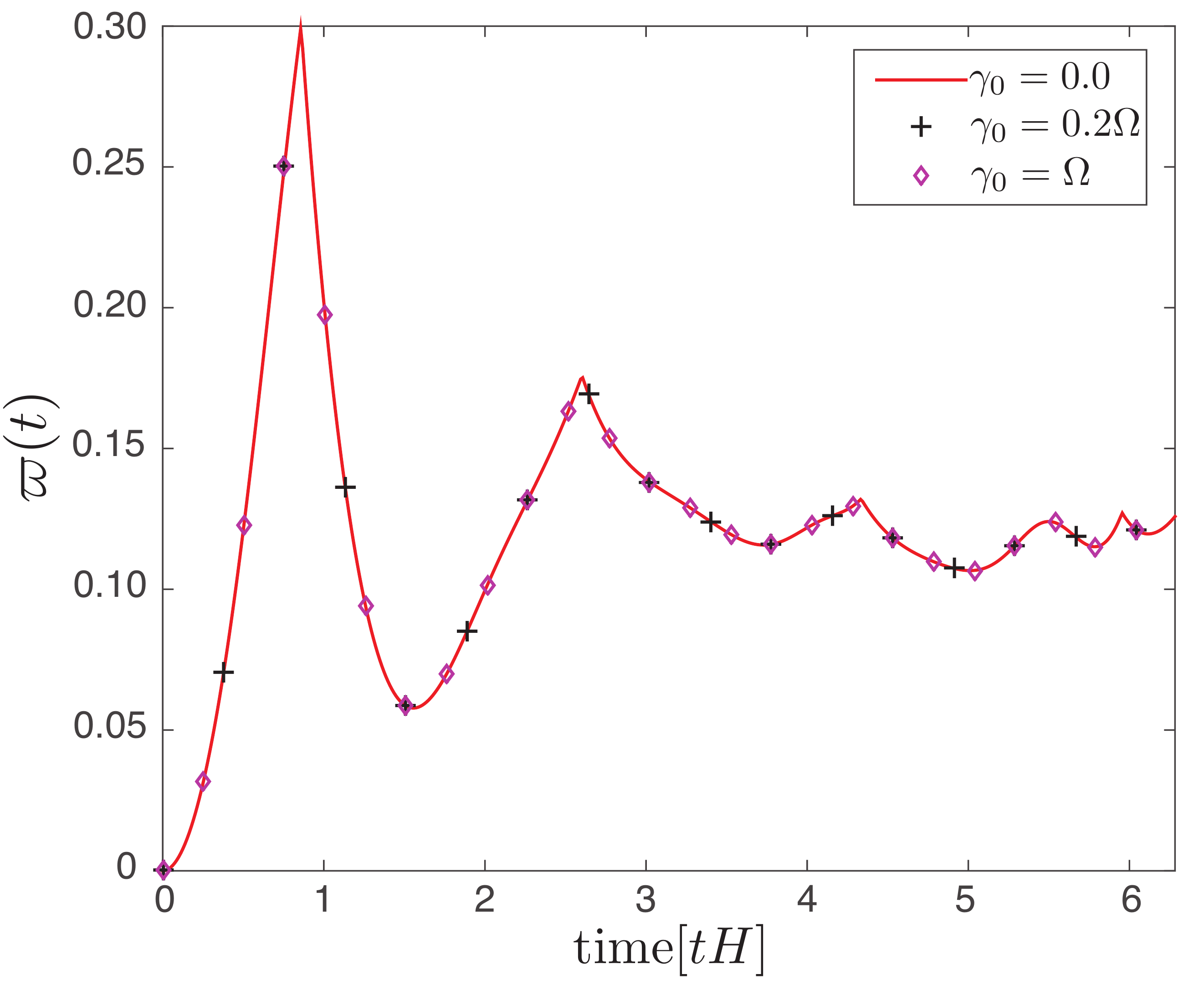}
\caption{Time-dependent rate function in one period $T$ is plotted for various decay rates associated with different Markovian baths. We have $N_A=6$ spins for the subsystem $A$ and $N_B=2$ spins for the subsystem $B$. The interspin coupling is $\tau=0.42\Omega$ and the transverse magnetic field $H=\Omega$. There is no current present in the subsystem $B$ initially, i.e., $\nu=0$ in Eq. (\ref{ring_B_Hamiltonian}). Thus, the entire ring does not couple to the environments as seen in a single return rate for three different Markovian decay rates.}
\label{fig:noCurrent_lambda}
\end{figure}

\begin{figure}[t]
\centering
\includegraphics[scale=0.4]{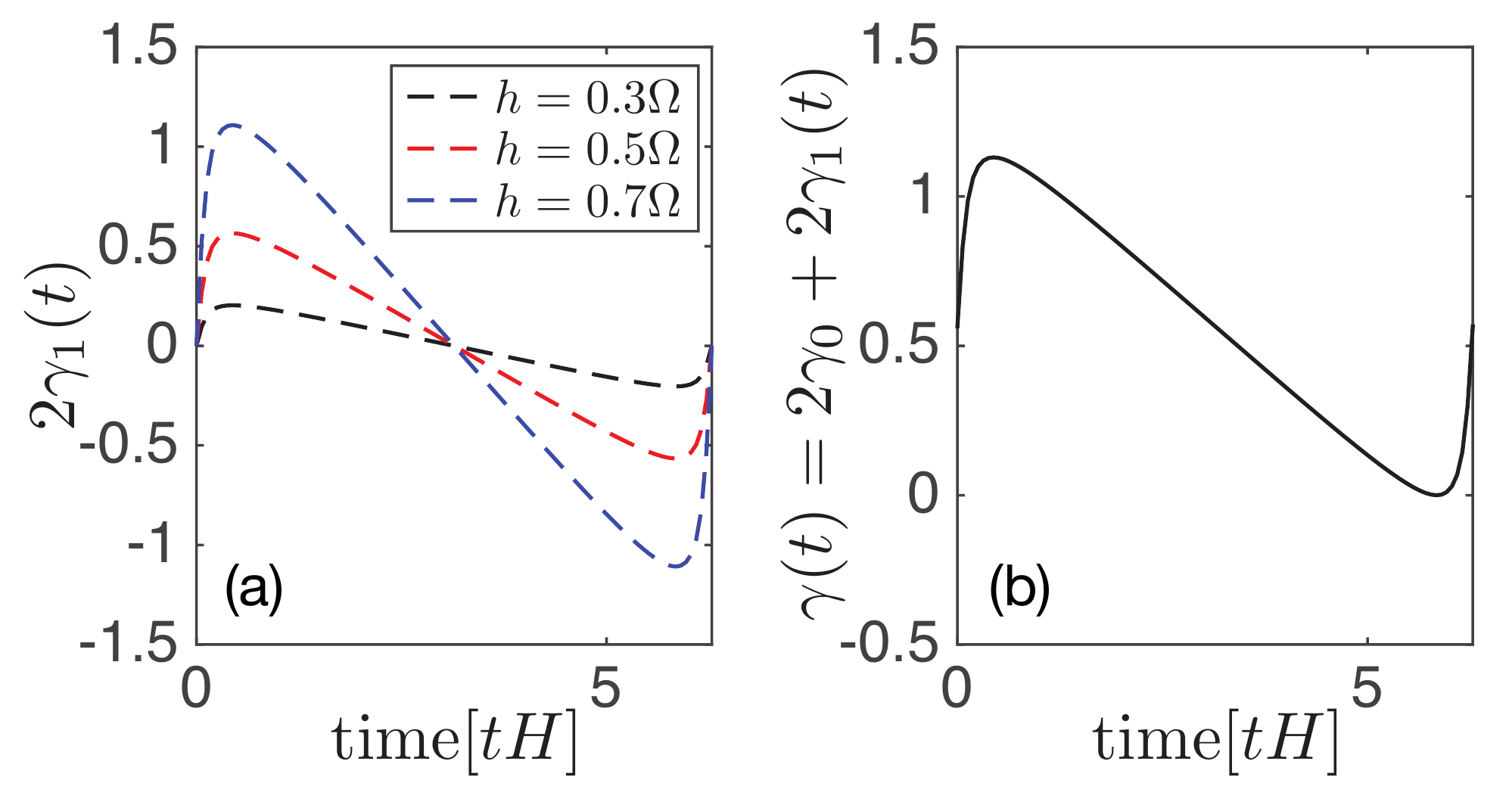}
\caption{(a) Various $\gamma_1 (t)$ versus time ($1T$) are plotted for three different values of $h=0.3 \Omega, 0.5 \Omega, 0.7 \Omega$. As seen from the figure, $\gamma_1$ exhibits both positive and negative values, with which the negative ones will contribute towards non-Markovian dynamics \cite{bastidas2018}. From the plots, we obtain the maximum values that are then used for Markovian dephasing rates in the upcoming numerical analyses. (b) Total $\gamma (t)$ is plotted in $1T$ for $\gamma_0 =\textrm{max}[\gamma_1(t)] = 0.2827\Omega$, and $h=0.5 \Omega$, such that the entire $\gamma(t)$ does not have negative value within one period. Hence, we are in complete Markovian evolution. We consider $M=60$ bosonic modes in the non-Markovian bath. See \ref{appen:general} for the open quantum system framework used.}
\label{fig:different_gamma}
\end{figure}

To derive the exact master equation, let us assume that the Ising ring couples to the baths via the global energy current $\hat{J}$, as shown in Fig.~\ref{fig:spinRing_baths}. We also consider an uncorrelated initial state $\hat{\rho}(0)=\hat{\rho}_S(0)\otimes\hat{\rho}_R(0)$, where $\hat{\rho}_S(0)$ and $\hat{\rho}_R(0)$ are the initial states of the system and reservoir, respectively. We quench the entire chain with the Hamiltonian of the total system
\begin{equation}
	\hat{H} = \hat{H}^S + \hat{J} \sum_{l}\hat{X}^l +\hat{H}_R .
\end{equation}
In this expression, the global coupling between the spin chain and the reservoir is defined in terms of the operator $\hat{X}^l=\hat{X}^l_{MB}+\hat{X}^l_{NMB}$, where $\hat{X}^l_{NMB}= g_{l} \left(\hat{b}_{l} ^\dagger + \hat{b}_{l} \right)$ and $\hat{X}^l_{MB}= \tilde{g}_{l} \left(\hat{c}_{l} ^\dagger + \hat{c}_{l} \right)$ are the quadratures of the Markovian ($MB$) and non-Markovian ($NMB$) baths, respectively. 
Similarly, $\hat{H}_R=\hat{H}_{MB}+\hat{H}_{NMB}$ is the Hamiltonian of the reservoir, where $\hat{H}_{MB}=\sum_l\tilde{\omega}_l\hat{c}_l^{\dagger}\hat{c}_l$ and $\hat{H}_{NMB}=\sum_l\omega_l\hat{b}_l^{\dagger}\hat{b}_l$. Note that the coupling to the dephasing baths is via a current operator $\hat{J}$ resembling the Dzyaloshinki-Moriya interaction in spin systems~\cite{dzyaloshinsky1958,moriya1960}.
In this manuscript, we consider the non-Markovian coupling strength $g_l=(h/\Omega^2)e^{-zl/2}$, where $z>0$ is a positive number, and the modes of the non-Markovian bath have frequencies $\omega_l=l\Omega$ ($l=1,2,...M$), where $\Omega$  is the fundamental frequency. Notice that the choice of units for $g_l$ is consistent with units of the energy current $\hat{J}$, so that the Ising ring-bath coupling has frequency in units such that $\hbar=1$.

\begin{figure}[t]
\centering
\includegraphics[scale=0.6]{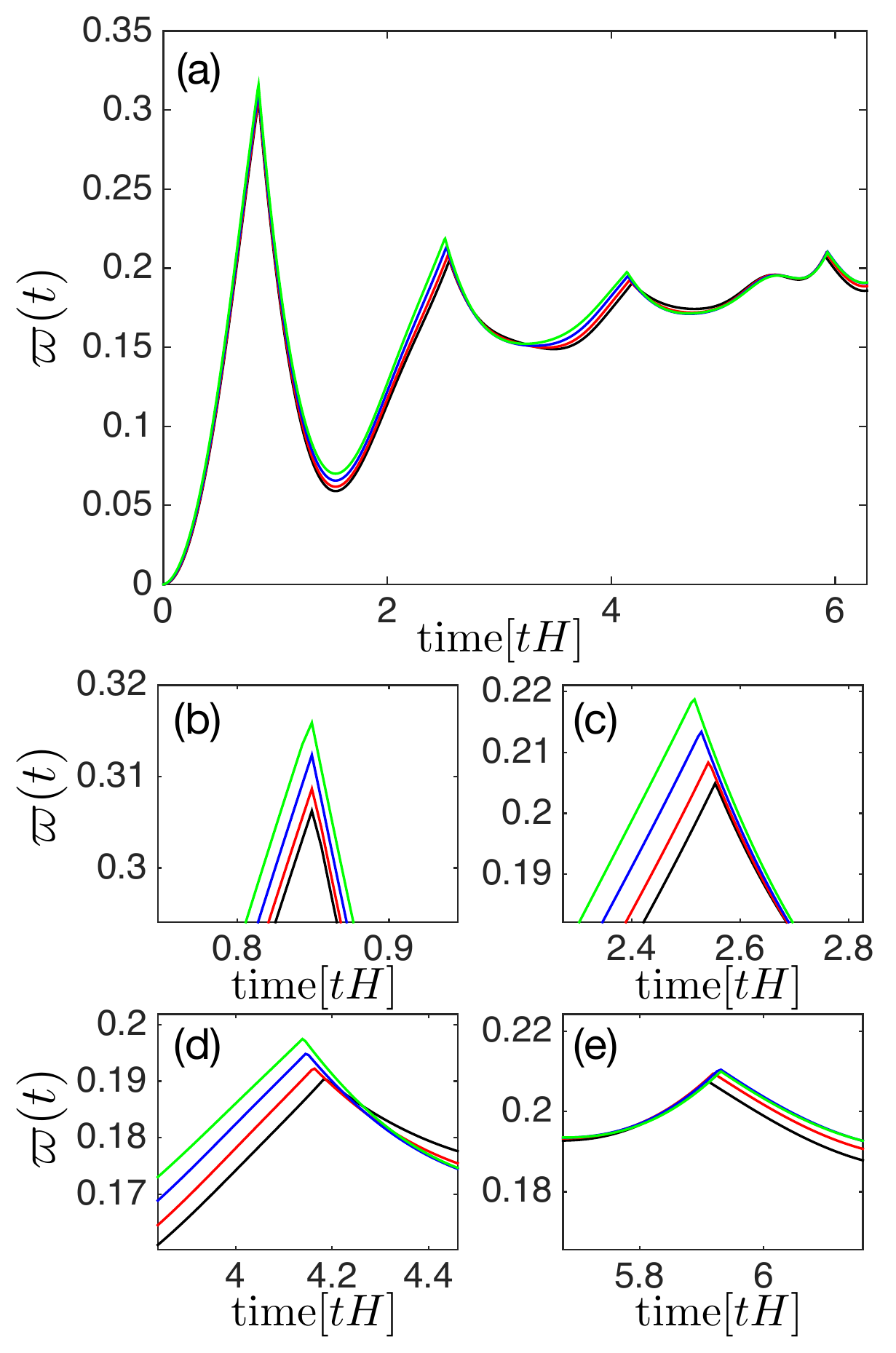}
\caption{(a) Time-dependent rate function in one period $T$ is plotted for various decay rates associated with different Markovian decay rates $\gamma_0$. Black line stands for $\gamma_0=0$, red line $\gamma_0=0.1018\Omega$, blue line  $\gamma_0=0.2827\Omega$, and green line  $\gamma_0=0.5542\Omega$.  (b-e) shows enlarged regions of four different critical times seen in (a). We have $N_A=6$ spins for the subsystem $A$ and $N_B=2$ spins for the subsystem $B$. The interspin coupling is $\tau=0.42\Omega$, the transverse magnetic field $H=\Omega$ and $\nu=5\Omega$. $h=0$ for all the plots and $z=0.1$. The number of bosonic modes in the non-Markovian bath is $M=60$.}
\label{fig:vary_gamma_oneT_combined}
\end{figure}
\begin{figure}[t]
\centering
\includegraphics[scale=0.6]{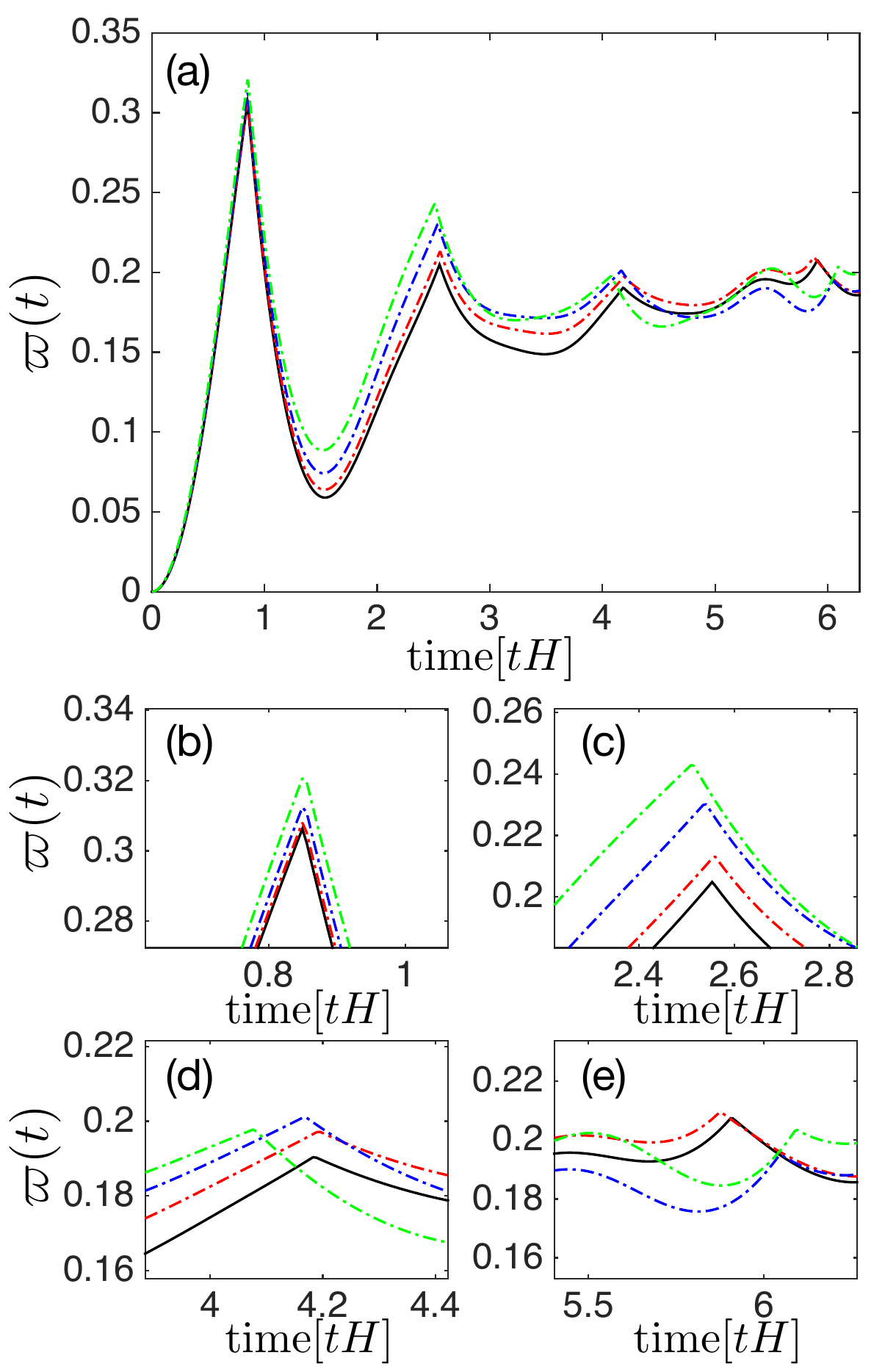}
\caption{(a) Time-dependent rate function in one period $T$ is plotted for various decay rates associated with different coupling strengths to the non-Markovian bath $h=0$ black line, $h=0.3\Omega$ red-dashed line, $h=0.5\Omega$ blue-dashed line, $h=0.7\Omega$ green-dashed line. (b-e) shows enlarged regions of four different critical times seen in (a). We choose the decay rates such that  $\textrm{max}[\gamma(t)]=\gamma_0$, where $\gamma_0$ are the parameters used in Fig. \ref{fig:vary_gamma_oneT_combined}. We have $N_A=6$ spins for the subsystem $A$ and $N_B=2$ spins for the subsystem $B$. The interspin coupling is $\tau=0.42\Omega$, the transverse magnetic field $H=\Omega$ and $\nu=5\Omega$. We choose a dephasing rate $\gamma(t)=\gamma_1(t)$ for all the plots and $z=0.1$ and the number of bosonic modes in the non-Markovian bath is $M=60$.}
\label{fig:vary_h_oneT_combined}
\end{figure}

According to the above description, the Ising ring evolves under the Lindblad-type master equation $d{\hat{\rho}}_S(t)/dt=-i[\hat{H}^{S} (t),\hat{\rho}_{S} (t)]+\gamma(t)\hat{\mathcal{D}}[\hat{J}] \hat{\rho}_S(t)$, where $\hat{\mathcal{D}}[\hat{J}]\hat{\rho}_{S} (t)=\hat{J}\hat{\rho}_{S} (t) \hat{J}- \frac{1}{2}\left(\hat{
J}^2 \hat{\rho}_{S} (t) + \hat{\rho}_{S} (t) \hat{J}^2 \right) $ and $\hat{H}^{\text{S}} (t) = \hat{H}^{\text{S}}-  \lambda(t) \hat{J}^2$, where there is a Lamb shift term proportional to
\begin{equation}
	\lambda(t) = \sum^M_{l=1} \frac{g^2_{l}}{\omega_{l}}  [1 -\cos (\omega_{l} t)].
\end{equation}
Also, the dephasing rate reads
\begin{equation}
	\gamma(t) =2\gamma_0+2\gamma_1(t),
\end{equation}
where $\gamma_0$ and $\gamma_1(t)=\sum^M_{l=1} \frac{g^2_{l}}{\omega_{l}}  \sin (\omega_{l} t) \coth \left(\beta_{NMB} \omega_{l}/2\right)$ are the dephasing rates associated with the Markovian and non-Markovian baths, respectively. Due to the choice of the frequencies of the bath $\omega_l=l\Omega$, the functions $\lambda(t)$ and $\gamma(t)$ are periodic with a period $T=2\pi/\Omega$. We also consider that the non-markovian is initially prepared in a thermal state with inverse temperature $\beta_{NMB}$. Note that in the previous equations, $M$ denotes the number of modes of the non-Markovian bath.
  It is worth mentioning that a crucial point in obtaining the above Lindblad equation is that both baths couple to a conserved quantity of the system, in our case, $[\hat{H}^S , \hat{J}]=0$. Thus, our theory may apply to any quantum system experiencing DQPT that couple to dephasing baths via global or local conserved quantities. See \ref{appen:general} and Ref.~\cite{bastidas2018} for a detailed explanation of our open quantum system theory. On important aspect of the master equation is that due to the time-periodic character of the rates, the dynamics of the Ising ring is divisible at stroboscopic times, which is referred to as Floquet stroboscopic divisibility~\cite{bastidas2018}.

\begin{figure}[t]
\centering
\includegraphics[scale=0.55]{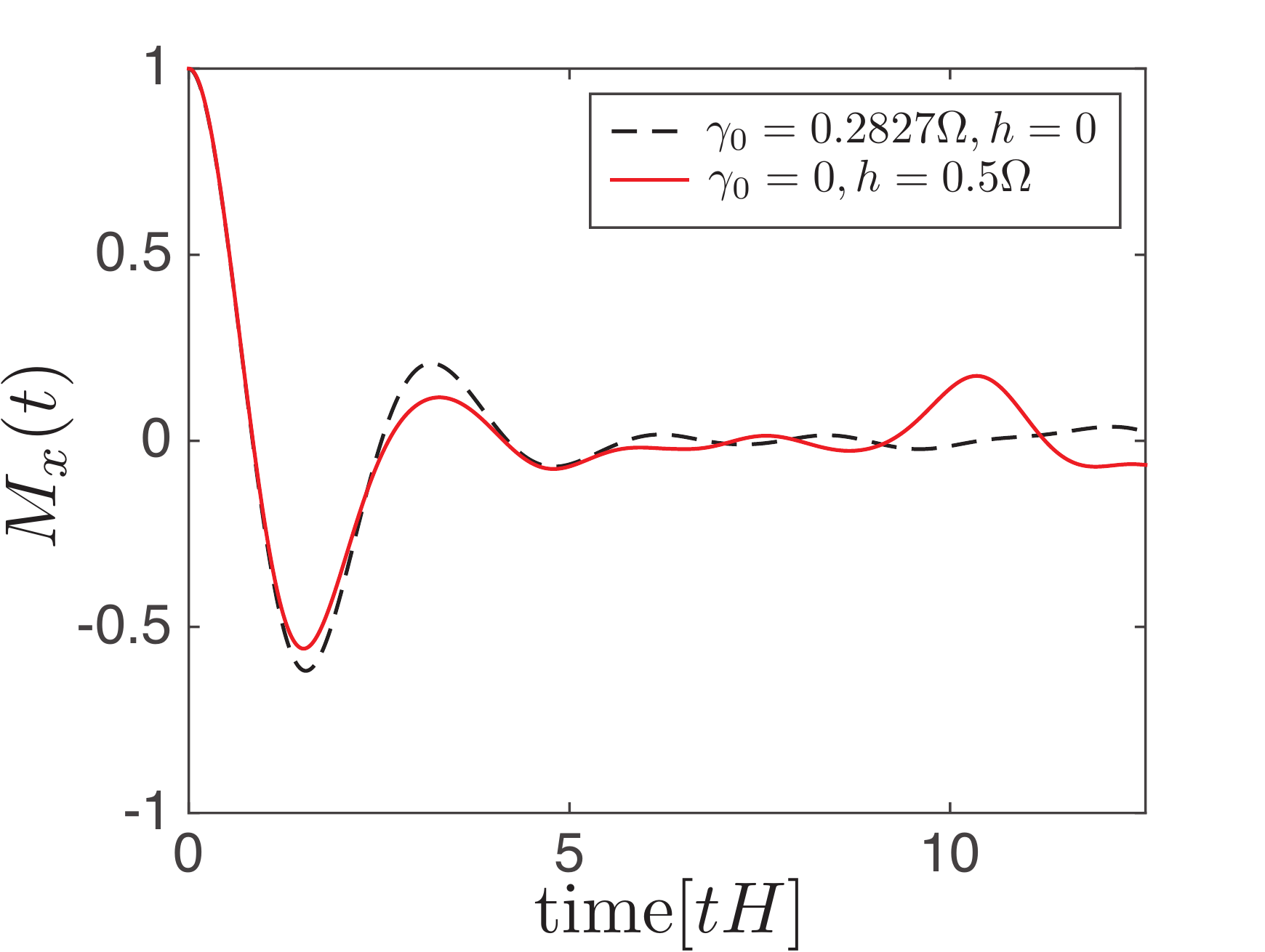}
\caption{The average magnetization in two periods $(2T)$ is plotted for purely Markovian dynamics (black solid line) and non-Markovian one (red solid line). The usual oscillatory decay of the order parameter is present for both cases, but non-zero oscillations are present in non-Markovian case while near zero order parameter is observed in the Markovian dynamics. In this simulation we have used the same interspin coupling, transverse magnetic field, and number of bosonic modes as in Fig.~\ref{fig:vary_h_oneT_combined}.}
\label{fig:markov_non-Markov_twoT}
\end{figure}


So far, we have derived the exact master equation governing the evolution of the Ising ring. In our manuscript, we focus on signatures of DQPTs that appear in the dynamics of the spin chain A, described by the reduced density matrix $\hat{\rho}_A(t)=\textrm{Tr}_B [\hat{\rho}_S(t)]$.  During the quench, the indirect signatures of the DQPTs for a finite system size \cite{jurcevic2017direct} can be obtained by measuring the two probabilities \cite{heyl2014dynamical} to return to the same initial ground state or the other (c.f. Eq. (\ref{eq:fidelity_Loschmidt}))-
$
	\mathcal{G}_{F,d}= \textrm{Tr}\left[\sqrt{\sqrt{\hat{\rho}_d}\hat{\rho}_A(t)\sqrt{\hat{\rho}_d}}\right],
$
where $d\in \{+,-\}$, $\hat{\rho}_+ =\ketbra{\psi_+}{\psi_+}$, $\hat{\rho}_- =\ketbra{\psi_-}{\psi_-}$. Inspired by a recent experiment with trapped ions~\cite{jurcevic2017direct}, we define the time-dependent rate function as 
\begin{equation}\label{main_rate_function}
	\varpi(t)\equiv \min_{d \in \{+,- \}} (-N^{-1}\log [\mathcal{G}_{F,d} (t)] ).
\end{equation}

First, we look at the simplest scenario where the system $A+B$ is completely isolated from environment. We numerically calculate the return rate $\varpi(t)$ and the expectation of the average magnetization along x-
direction,
$
	M_x = \langle \hat{\mathcal{M}}_x \rangle=\langle N^{-1}\sum_i \hat{\sigma}_i ^x \rangle
$. The latter is the order parameter for the Ising chain $A$. We also plot the probabilities $P_\pm (t)$ to return to the states $\hat{\rho}_{\pm}$, where 
$
P_\pm (t) = \textrm{Tr}[\hat{\rho}_\pm . \hat{\rho}_A (t)]/P,
$ 
and $P=\textrm{Tr}[\hat{\rho}_+ .\hat{\rho}_A (t)]+\textrm{Tr}[\hat{\rho}_- .\hat{\rho}_A (t)]$.

For proper comparison, in all the numerics, we set $\nu=5\Omega$, and we quench with the Hamiltonian $\hat{H}^{S}$, Eq. (\ref{eq:HamIsing}), with $\tau=0.42\Omega$ and $H=\Omega$. The results are shown in Fig. \ref{fig:no_bath_collage}, which is in agreement with existing literature \cite{heyl2014dynamical,jurcevic2017direct}. The sign of magnetization $M_x$ changes at the critical critical times $t_c$, where the two probabilities $P_+$ and $P_-$ cross, eventually leading to an oscillatory decay of the expectation value $\langle \hat{\mathcal{M}}_x \rangle$. This is the clear evidence of the DPT-II \cite{vzunkovivc2018dynamical,heyl2014dynamical}. 

\subsection{Numerical signatures of the DQPTs in an open quantum system}\label{subsec:signatures}

Before we discuss the numerical evidence of DQPTs in the open quantum system, we would like to remark why we introduce the energy current inside the system via Eq. (\ref{ring_B_Hamiltonian}). The main reason is that our system couples to baths via the conserved quantity called the energy current density $\hat{J}$ (\ref{global_current}), as prescribed above. As a consequence, if there is no current present inside the Ising ring, the dephasing channels plays no role, independent on how strongly the system couples to the baths. The system is coupled to the environment when the initial total current is different from zero, i.e., $\langle\hat{J}(0)\rangle=\langle\hat{J}_{A}(0)\rangle  +\langle\hat{J}_{B}(0) \rangle \neq 0$. For simplicity, let us consider the Ising ring coupled to a Markovian bath. Numerical evidence from Fig. \ref{fig:noCurrent_lambda}, where we plot the rate function $\varpi$ for three different decay rates $\gamma_0$, clearly supports our discussion. To overcome this, we introduce the current $\langle\hat{J}_{B}(0) \rangle \neq 0$ inside the subsystem $B$ at the start of the quench.

Furthermore, our open quantum system approach (see \ref{appen:general}) allows us to tune the system-bath interactions to discuss different scenarios involving:  i) only a Markovian environment, ii) only non-Markovian bath \& iii) the combined effect of both Markovian and non-Markovian baths. In this way, we can study the effect of system-bath coupling on the dynamics of DQPTs. The rate $\gamma_1(t)$ is a periodic function, whose average is zero in one period, as seen in Fig. \ref{fig:different_gamma} (a). The dephasing rate $\gamma_0$ controls the overall sign of the dephasing rate $\gamma(t)$, since it acts like a dc signal. By tuning the dephasing rate $\gamma_0$ one can control if the dynamics is Markovian ($\gamma(t)\geq0$ for all times $t>0$) or non-Markovian when the rate becomes negative at certain time intervals. The time-dependence of the dephasing rate is depicted in Fig.~\ref{fig:different_gamma} (b). For a more detailed discussion of non-Markovian dynamics in master equations with time-periodic rates, we refer the reader to Ref.~\cite{bastidas2018}. In all our numerical calculations we set the parameters in terms of the fundamental frequency $\Omega$ that defines the period $T=2\pi/\Omega$ of the dephasing rate $\gamma(t+T)=\gamma(t)$.

In Fig.~\ref{fig:vary_gamma_oneT_combined}, we show numerical evidence of the DQPTs for various Markovian dephasing rates, in terms of the time-dependent rate function $\varpi (t)$. We show the enlarged regions of the four particular critical times seen in Fig. \ref{fig:vary_gamma_oneT_combined} (a) in the order of appearance in Fig. \ref{fig:vary_gamma_oneT_combined} (b-e). It is evident that when the dephasing rate increases, we see the shift in the dynamical critical times towards the left as compared to the closed system case (the black solid line), in particular (c) and (d).

After discussing the Markovian case, now we explore signatures of DQPTs in the case of non-Markovian dynamics. To do this, we consider $\gamma(t)=\gamma_1(t)$ and choose the coupling parameter $h$ in such a way that its maximum value $\textrm{max}[\gamma(t)]=\gamma_0$ will correspond to values of $\gamma_0$ depicted in Fig.~\ref{fig:vary_gamma_oneT_combined}. In this figure one can observe that the phase shift towards the left can still be seen in the non-Markovian ones for the second and third critical times. However, we observe the right shift at the last peak with the increase in the system-bath coupling strengths. The results are shown in Fig. \ref{fig:vary_h_oneT_combined} (a-e).

\subsection{Interplay between non-Markovian and Markovian environments}\label{subsec:interplay}

As seen in both Figs. \ref{fig:vary_gamma_oneT_combined} and \ref{fig:vary_h_oneT_combined}, the different effects of both Markovian and non-Markovian baths on the DQPTs signatures are not obvious when we look at the rate function $\varpi (t)$. However, when we consider an order parameter such as the average magnetization $M_x$ for the two different cases (see Fig.~\ref{fig:markov_non-Markov_twoT}), we observe the usual oscillatory decay in both cases. However, the oscillations are more pronounced in the non-Markovian case, even at time scales where the oscillations disappear under the effect of a Markovian bath.

\begin{figure}[t]
\centering
\includegraphics[scale=0.44]{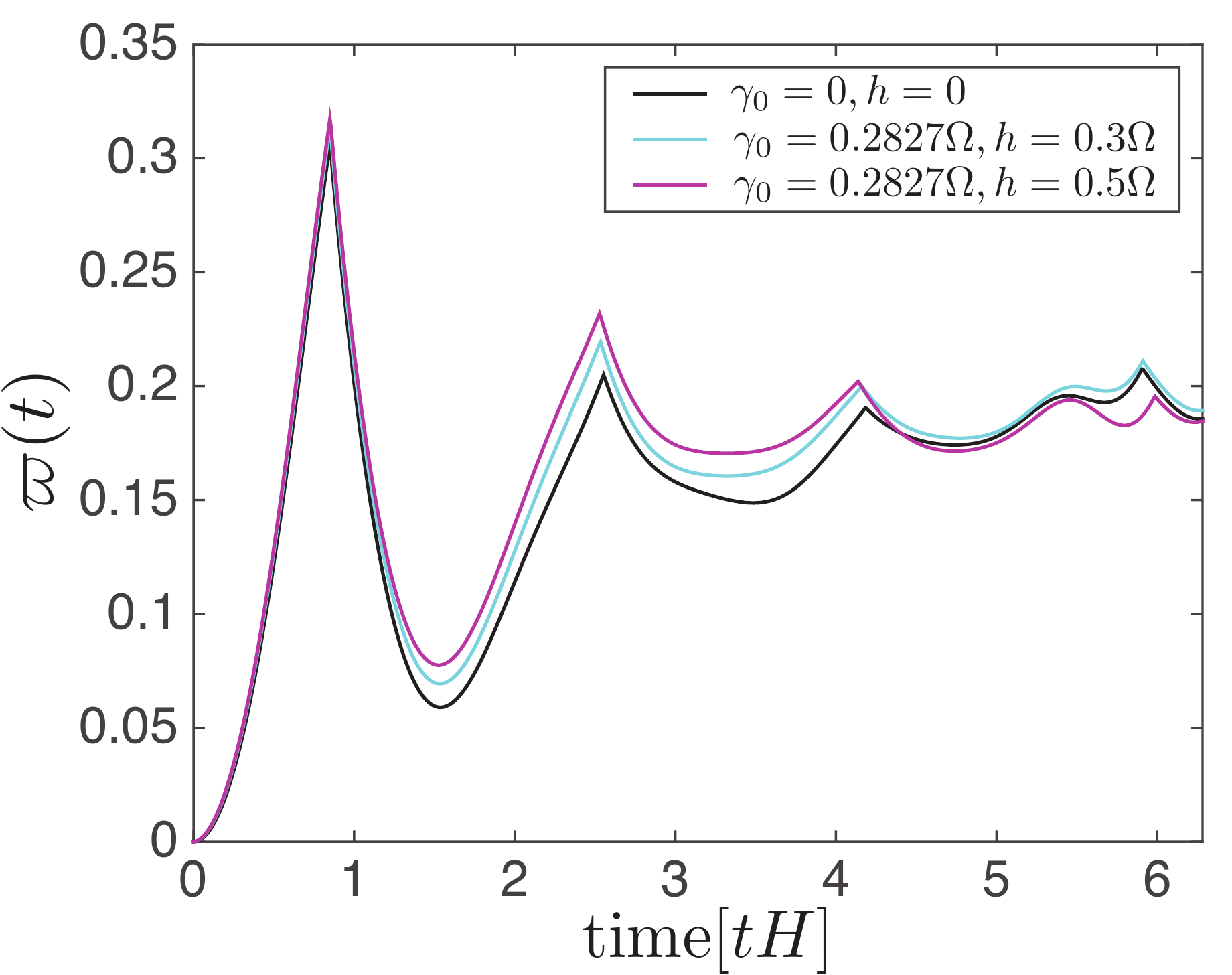}
\caption{Time-dependent rate function in one period $T$ is plotted for various decay rates associated with different Markovian and non-Markovian dephasing rates, $\gamma_0$ and $h$. In this simulation we have used the same interspin coupling, transverse magnetic field, and number of bosonic modes as in Fig.~\ref{fig:vary_h_oneT_combined}.}
\label{fig:fig9}
\end{figure}

\begin{figure}[t]
\centering
\includegraphics[scale=0.44]{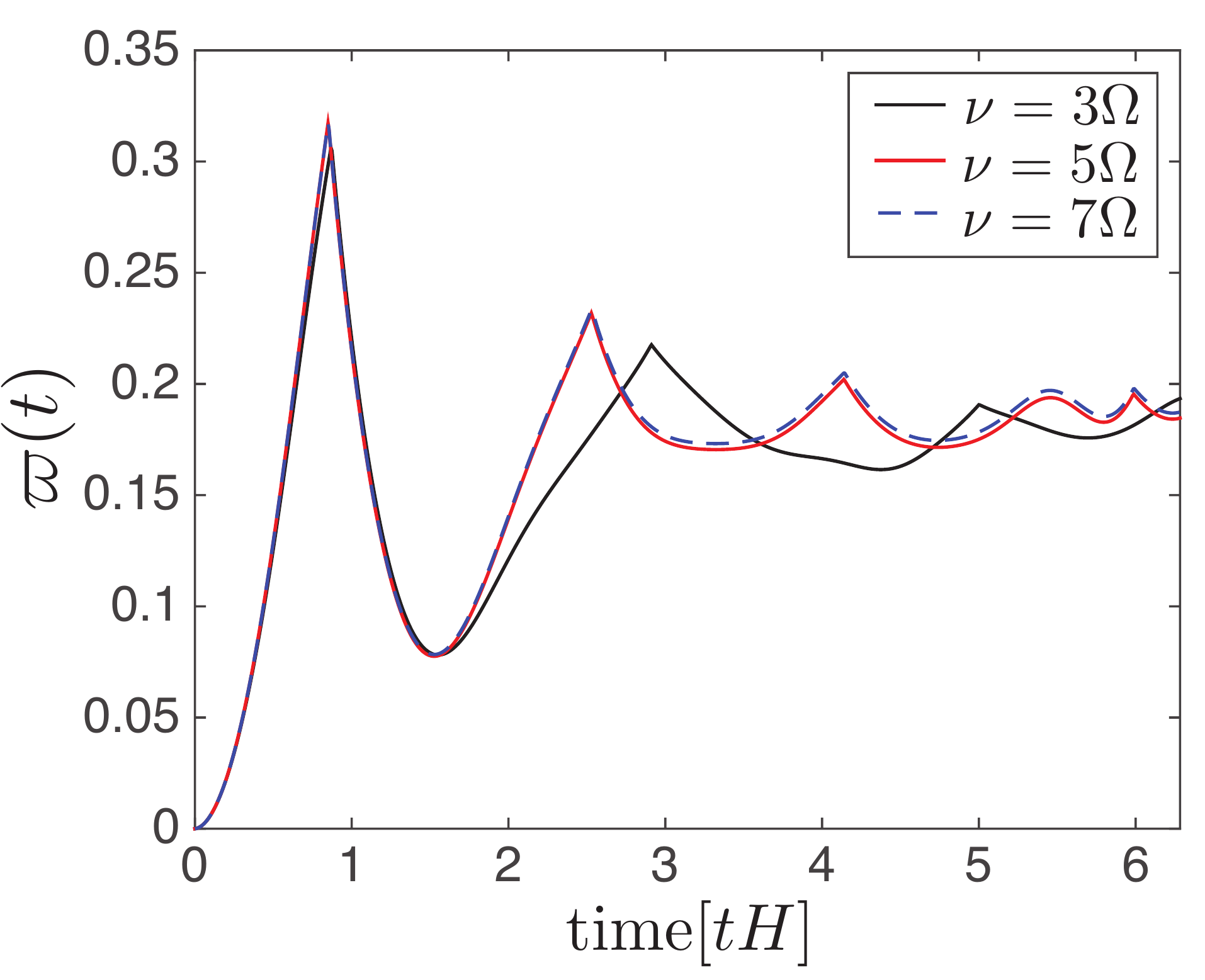}
\caption{Time-dependent rate function in one period $T$ is plotted for various intensities of the initial current which is controlled by the parameter $\nu$ in Eq.~(\ref{ring_B_Hamiltonian}). In this case, we consider the action of non-Markovian dephasing bath with parameters $h=0.5\Omega$, $z=0.1$, and $60$ modes, and for the Markovian bath with use $\gamma_0=0.2827\Omega$. The parameters for the Ising ring are $H=\Omega$ and $\tau=0.42\Omega$.}
\label{fig:fig10}
\end{figure}

For better comparison, in Fig.~\ref{fig:fig9} we plot the rate function $\varpi (t)$ for both Markovian and non-Markovian dephasing baths. Also, in Fig.~\ref{fig:fig10} we show the rate function for different values of the initial current in the Ising ring, which is controlled by the parameter $\nu$ in Eq.~(\ref{ring_B_Hamiltonian}). From these plot it is clear that no matter how strong the initial current in the ring is, the position of the first critical point showing signatures of DQPTs remains invariant. This a remarkable result, that is, DQPT in our Ising ring scheme is robust under the action of both Markovian and non-Markovian dephasing baths at short time scales. 

\section{Conclusion and outlook}\label{sec:conclusion}
In summary, we have presented a theoretical framework to investigate dynamical quantum phase transitions under the action of Markovian and non-Markovian dephasing baths. We have shown that with the help of the Floquet stroboscopic dynamics (see \ref{appen:general}), one can tune the Markovian and non-Markovian dynamics to observe their effects on the DQPTs signatures. In particular, we have studied the paradigmatic transverse field Ising model with periodic boundary condition. In this situation, the energy current is a conserved quantity, which we use to couple the Ising ring to Markovian and non-Markovian dephasing baths. We have demonstrated that no matter how intense the initial current in the ring is, the positions of critical points showing signatures of DQPTs remain invariant. In this case, signatures of DQPT in our Ising ring scheme are robust under the action of both dephasing baths.  We believe our result would open up a new window of opportunities in the direction of non-equilibrium quantum phase transitions in the open quantum systems with the strong system-bath coupling.

\section*{Acknowledgement}
The authors are grateful for the financial support through the National Research Foundation and the Ministry of Education Singapore. G.R. acknowledges the support from the Fondo Nacional de Desarrollo Cient\'ifico y Tecnol\'ogico (FONDECYT, Chile) under grant No. 1190727.

\appendix

\section{Derivation of the conserved quantity: the global current}
\label{appen:global_current}

A general flux (current) operator may be obtained by assuming that there exists an operator
continuity equation in one dimension \cite{wu2008energy}
\begin{align}
\frac{\partial \hat{h}(x,t)}{\partial t} + \frac{\partial \hat{j}(x,t)}{\partial x}=0,
\end{align}
where $\hat{h}(x,t)$ and $\hat{j}(x,t)$ are the energy density and the energy flux operators respectively.
For an $N$-site chain with $m$ states at each site, one can write down the energy density operator as such
\begin{align}
\hat{h}(x,t)=\sum_s \hat{h}_s\delta(x-x_s),
\end{align}
with $\hat{h}_s$ being the discrete energy operator at the site $s$, and $\delta(x-x_s)$ is the Dirac delta function, so that the system
Hamiltonian reads $\hat{H}=\int dx~\hat{h}(x,t)=\sum_s \hat{h}_s$. Similarly, the energy
flux operator is
\begin{align}
\hat{j}(x,t)=\sum_s\hat{j}_s\delta(x-x_s),
\end{align}
where $\hat{j}_s$ is the current operator at the $s$th site. Under these definitions, the 
continuity equation adopts the following discrete form
\begin{eqnarray}
\frac{\partial \hat{h}(x,t)}{\partial t}&=&\sum_s\frac{d\hat{h}_s}{dt}\delta(x-x_s),\\
\frac{\partial \hat{j}(x,t)}{\partial x}&=&\sum_s -\bigg(\frac{\hat{j}_{s-1}-\hat{j}_s}{a}\bigg)\delta(x-x_s),\\
\frac{d\hat{h}_s}{dt}&=&\frac{\hat{j}_{s-1}-\hat{j}_s}{a},
\label{current_conservation}
\end{eqnarray}
where $a$ is the spacing between any $j+1$th and $j$th sites. The time evolution of the operator $\hat{h}_s$ in the 
Heisenberg picture reads
\begin{align}
\frac{d\hat{h}_s}{dt}=i[\hat{H},\hat{h}_s].
\end{align}
To proceed further, one may consider a generic 1D Hamiltonian with up to two-body nearest-neighbor
interactions,
\begin{align}
\hat{H} = \sum_s(\hat{h}^0_s+\hat{V}(s,s+1)),
\end{align}
where $\hat{h}^0_s$ is the local Hamiltonian at site $s$, and $\hat{V}(s,s+1)$ is the site-dependent two-body interaction terms. When we invoke this discretisation formalism to the Ising ring model (see Fig. \ref{fig:closedRing}), we arrive at the global current operator \cite{antal1997}
\begin{equation}
	\hat{J} = \frac{H\tau}{2}\sum_j \hat{\sigma}^y _j \left(\hat{\sigma}^x _{j-1}- \hat{\sigma}^x _{j+1} \right).\nonumber
\end{equation} 
We note that the terms inside the above summation is the Dzyaloshinsky-Moriya interaction in the theory of weak ferromagnetism \cite{dzyaloshinsky1958,moriya1960}. One direct consequence is that we have a conserved quantity $\hat{J}$ such that $[\hat{H}^S,\hat{J}]=0$ when we have a close boundary condition, i.e., $\hat{\sigma}_{N+1}=\hat{\sigma}_1$.

\section{A roadmap to open quantum systems}\label{appen:general}
\begin{figure}[t]
\centering
\includegraphics[scale=0.5]{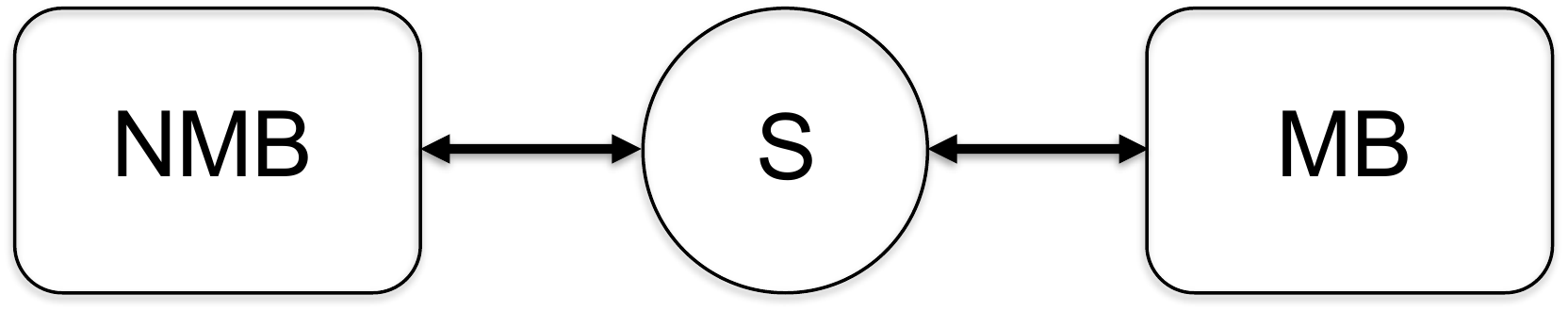}
\caption{An arbitrary quantum system (S) interacts with both Markovian and non-Markovian baths labelled as 'MB' and 'NMB' respectively.}
\label{fig:general_open_quantum}
\end{figure}

Let us denote a system Hamiltonian as $\hat{H}_S$, where it can, in principle, be quantum Ising chain, atom-atom interaction in ultracold atoms lattices, Bose-Hubbard model, etc. The environment degrees of freedom are captured by bosonic harmonic oscillators. We denote it as $\hat{H}_R$, which is composed of $\hat{H}_{NMB}= \sum_{j} \omega_{j} \hat{b}_{j}^\dagger \hat{b}_{j}$ and $\hat{H}_{MB}= \sum_{j} \tilde{\omega}_{j} \hat{c}_{j}^\dagger \hat{c}_{j}$. $\hat{b} (\hat{b}^\dagger)$ are bosonic annihilation (creation) operators for non-Markovian bath, while $\hat{c} (\hat{c}^\dagger)$ are bosonic annihilation (creation) operators for Markovian one. The setup is that the system described by $\hat{H}_S$ couples globally to the environment $\hat{H}_B$ via a conserved quantity as depicted in Fig. \ref{fig:general_open_quantum}. A microscopic derivation of the reduced system dynamics is obtained by considering the following total Hamiltonian
\begin{equation}
	\hat{H} = \hat{H}_S + \hat{V}_S \sum_{l} \hat{X}^l+\hat{H}_R ,
\end{equation}
where $\hat{V}_S$ is the system operator and $\hat{X}^l=\hat{X}^l_{MB}+\hat{X}^l_{NMB}$. For simplicity, we choose $\hat{X}^l_{NMB}$ and $\hat{X}^l_{MB}$ to be of the form $\kappa_{l} \left(\hat{\Theta}_{l} ^\dagger + \hat{\Theta}_{l} \right)$. We remark that $\kappa_{l}=g_{l}$ represents the coupling between the system operator and the bosonic mode $\omega_j$ of the non-Markovian $(\hat{\Theta}=\hat{b})$ and $\kappa_{l}=\tilde{g}_l$ is the Markovian coupling with $\tilde{\omega}_l$ bosonic modes $(\hat{\Theta}=\hat{c})$. $M$ is the total number of modes inside each reservoir. Although the system-bath coupling is treated globally, it is noteworthy that our previous work \cite{bastidas2018} also applies to the local system-bath coupling. The only constraint we impose to derive our microscopic master equation reads
\begin{equation}
      \label{eq:LocalConsQuant}
	[\hat{H}_{S}, \hat{V}_S]=0.
\end{equation}

Whenever the above constraint, Eq. (\ref{eq:LocalConsQuant}), is satisfied, we can always find the reduced density matrix of the system analytically \cite{bastidas2018}, and it is found to be  
\begin{equation}
       \label{eq:RedSystDenMat}
	\hat{\rho}_S(t)=\text{Tr}_B[\hat{\rho}(t)]=\sum_{\alpha,\beta} c_\alpha c_\beta ^\ast e^{-i(E_\alpha -E_\beta) t}F_{\alpha\beta}(t)\ketbra{E_\alpha}{E_\beta} ,
\end{equation}
where $c_\alpha$ are complex coefficients in the system eigenbases $\ket{E_\alpha}$ with eigenenergies $E_\alpha$. We have assumed that the combined system and bath are initialized in a product state $\hat{\rho}(0)=\hat{\rho}_S(0)\otimes\hat{\rho}_B(0)$, where
$\hat{\rho}_S(0)=\sum_{\alpha,\beta} c_\alpha c_\beta ^\ast \ketbra{E_\alpha}{E_\beta}$ and $\hat{\rho}_B(0)$ are the system and bath initial density matrices. In addition, we assume the two baths initially to be $\hat{\rho}_B(0)=\hat{\rho}_{MB}(0)\otimes\hat{\rho}_{NMB}(0)$ a product of thermal states with inverse temperatures $\beta_{MB}$ and $\beta_{NMB}$, respectively.
This means that $\hat{\rho}_{a}(0)=e^{-\beta_a\hat{H}_a}/Z_a$, where $Z_a=\text{Tr}[e^{-\beta_a\hat{H}_a}]$ with $a\in\{MB,NMB\}$. Moreover,
\begin{align}
      \label{eq:InfFunc}
	   F_{\alpha\beta}(t)&=e^{-\gamma^{0} t\left(V^{(\alpha)}-V^{(\beta)}\right)^2} \text{Tr}\left[\hat{\rho}_{NMB}(0) e^{i\hat{H}^{(\beta)}_{SB}t} e^{-i\hat{H}^{(\alpha)}_{SB}t}\right]
	   \nonumber \\
	   &=e^{-\Gamma(t)\left(V^{(\alpha)}-V^{(\beta)}\right)^2 +i\Lambda(t)\left\{\left[V^{(\alpha)}\right]^2-\left[V^{(\beta)}\right]^2\right\}} 
\end{align}
is the time-dependent influence functional that dictates the incoherent processes of the reduced system dynamics, with the functions
\begin{align}
      \label{eq:ParametersInfluenceFunc}
            \Lambda(t) &= \sum_{l} \left(\frac{g_{l}}{\omega_{l}} \right)^2 [\omega_{l} t -\sin (\omega_{l} t)], \text{  \&}\\
	\Gamma (t) &= \gamma^{0} t+\sum_{l} \left(\frac{g_{l}}{\omega_{l}} \right)^2 [1-\cos (\omega_{l} t)] \coth \left(\frac{\beta_{NMB} \omega_{l}}{2}\right)
	\ .
\end{align}
The dephasing rate $\gamma^{0}$ is due to the coupling to the Markovian bath.
With the constraint Eq.~\eqref{eq:LocalConsQuant}, $\hat{H}_S$ and $\hat{V}_S$ can be simultaneously diagonalised in the same bases, i.e., $\hat{V}_S\ket{E_\alpha}=V^{(\alpha)}_S\ket{E_\alpha}$, where $V^{(\alpha)}_S$ denotes the eigenvalue of the operator $\hat{V}_S$, and $\ket{E_\alpha}$ are eigenvectors of the Hamiltonian $\hat{H}_S$. The operator $\hat{H}^{(\alpha)}_{SB}$ appearing in Eq. \eqref{eq:InfFunc} reads $\hat{H}^{(\alpha)}_{SB}=V^{(\alpha)}_S \sum_{l} g_{l} \left(\hat{b}_{l}^\dagger + \hat{b}_{l} \right)$. We note that for each eigenstate $\ket{E_\alpha}$, the Hamiltonian  $\hat{H}^{(\alpha)}_{SB}$ describes a set of displaced bosonic harmonic oscillators whose displacement is proportional to the quantity $g_{l}V^{(\alpha)}_S/\omega_{l}$ \cite{bastidas2018}.

When we take the time derivative of the exact solution for $\hat{\rho}_S(t)$ in Eq.~\eqref{eq:RedSystDenMat}, we arrive at 
\begin{equation}
      \label{eq:SysMasterEq}
	\frac{d\hat{\rho}_{S} (t)}{dt}=\hat{\mathcal{L}}(t)\hat{\rho}_S (t)
	= -i[\hat{H}^{S} (t),\hat{\rho}_{S} (t)]+ \gamma(t) \hat{\mathcal{D}}[\hat{V}_S]\hat{\rho}_{S} (t),
\end{equation}
where the Lindblad superoperator is $\hat{\mathcal{D}}[\hat{V}_S]\hat{\rho}_{S} (t)=\hat{V}_S\hat{\rho}_{S} (t) \hat{V}_S- \frac{1}{2}\left(\hat{V}_S^2 \hat{\rho}_{S} (t) + \hat{\rho}_{S} (t) \hat{V}_S^2 \right) $ and $\hat{H}^{\text{S}} (t) = \hat{H}^{\text{S}}-  \lambda(t) \hat{V}_S^2$, where the Lamb shift term is proportional to
\begin{equation}
	\lambda(t) = \dot{\Lambda}(t)=\sum_{l} \frac{g^2_{l}}{\omega_{l}}  [1 -\cos (\omega_{l} t)] 
      \ ,
\end{equation}
where $\dot{\Lambda}$ is the time derivative of $\Lambda(t)$. Also, the dephasing rate reads
\begin{equation}
      \label{eq:nonLinDephasingRate}
	\gamma(t) = 2\dot{\Gamma}(t)
	=2\gamma_0+2\sum_{l} \frac{g^2_{l}}{\omega_{l}}  \sin (\omega_{l} t) \coth \left(\beta_{NMB} \omega_{l}/2\right)
	\ .
\end{equation}

We have so far derived a time-local Lindblad master equation, Eq.~\eqref{eq:SysMasterEq}, with time-dependent rates $\gamma (t)$, composed of $\gamma_0$ (the Markovian contribution) and $\gamma_1 (t)$ (the non-Markovian one comes from the second term in the summation (\ref{eq:nonLinDephasingRate})). It is commonly accepted that when one of the dissipation rates $\gamma(t)$ becomes negative at certain time interval, the dynamics is referred to as non-Markovian \cite{rivas2014quantum,breuer2016colloquium,de2017dynamics}. In contrast, if the rates are positive at all times the evolution is considered Markovian. In the non-Markovian regime, there is back-flow of information between the system and environment \cite{rivas2014quantum,breuer2016colloquium,de2017dynamics}. This means that when the rates are positive, the bath destroys coherent properties of the system. Contrary to this, when the rates are negative, the bath restores the lost information partially. 

In order to involve the floquet stroboscopic divisibility, we impose periodicity $\hat{\mathcal{L}}(t+T)=\hat{\mathcal{L}}(t)$ in the time-local master equation we obtain above $\frac{d\hat{\rho}_S (t)}{dt}=\hat{\mathcal{L}}(t)\hat{\rho}_S (t)$, where $\hat{\mathcal{L}}$ is a time-periodic Liouvillian operator (LO). This is possible if we choose the frequencies of the non-Markovian bath to be $\omega_{j}=j\Omega$ with $j$ being the integer multiples. The fundamental frequency $\Omega$, determines the period $T=2\pi/\Omega$ of the Liouvillian. Therefore, all the interesting properties of the Floquet theory can be directly applied \cite{bastidas2018}. For instance, one can define a propagator $\hat{\Phi}(t;0)$, or dynamical map, such that $\hat{\rho}_{{S}}(t)=\hat{\Phi}(t;0)\hat{\rho}_{{S}}(0)$. Due to the periodic nature of the LO, the dynamical map is divisible at stroboscopic times and  we have $\hat{\Phi}(mT;0)=[\hat{\Phi}(T;0)]^m$. Thus, one can focus the system dynamics just for one time period, due to the periodic nature of the master equation. Furthermore, we assume that the non-Markovian bath is prepared initially at near zero temperature, i.e., $\beta_{NM}\rightarrow\infty$ and consider couplings $g_{j}=(h/\Omega^2)e^{-zj/2}$, where $z>0$ is a positive number. In this way, one can couple an arbitrary quantum system with both Markovian and non-Markovian environments. Furthermore, one is able to tune the system-bath couplings explicitly, which is useful for the quantum bath engineering. In the main text, we have corroborated the general idea developed here with a standard example of quantum Ising spin chain in a periodic boundary condition, and look for the DQPTs signatures via the fidelity Loschmidt amplitude.

\section{Quench protocol}
Our quench protocol for open quantum system simulation is summarized in the table below. `ground' is shortened as `gnd.'
\begin{center}
 \begin{tabular}{||c| c| c| c| c||} 
 \hline
 Quench & Subsystem $A$ & Subsystem $B$ & Bath & System-bath \\ [0.5ex] 
 \hline\hline
 Before & one of the gnd. states$(\hat{H}^{S_1}_A)$ & gnd. state$(\hat{H}^{S_1}_B + \hat{H}^{S_2}_B -\nu \hat{J}_B )$ & nil & nil \\ 
 \hline
 During &  $\hat{H}^{S_1}_A + \hat{H}^{S_2}_A$ & $\hat{H}^{S_1}_B + \hat{H}^{S_2}_B$ & $\hat{H}_{R}$ & $\hat{J}\sum_l \hat{X}^l$ \\
 \hline
\end{tabular}
\end{center}

\section*{References}
\bibliographystyle{iopart-num}
\bibliography{Mybib}

\end{document}